\documentclass[%
 reprint,
 superscriptaddress,
 preprintnumbers,
 amsmath,amssymb,
 aps,
floatfix,
]{revtex4-2}

\usepackage[utf8]{inputenc}
\usepackage[T1]{fontenc} 
\usepackage{graphicx}
\usepackage{dcolumn}
\usepackage{bm}
\usepackage{xcolor}

\renewcommand{\arraystretch}{1.2}
\makeatletter
\newcommand{\mathleft}{\@fleqntrue\@mathmargin0pt}
\newcommand{\mathcenter}{\@fleqnfalse}
\makeatother

\begin{document}

\preprint{ADP-21-16/T1163}
\preprint{DESY 21-166}
\preprint{LTH 1270}

\title{State mixing and masses of the $\pi^0$, $\eta$ and $\eta^\prime$ mesons from $n_f=1+1+1$ lattice QCD+QED}

\author{Z.R.~Kordov}
\affiliation{%
 CSSM, Department of Physics, University of Adelaide, SA, Australia
}%
\author{R.~Horsley}
\affiliation{School of Physics and Astronomy, University of Edinburgh, Edinburgh EH9 3FD, UK}
\author{W.~Kamleh}%
\affiliation{%
 CSSM, Department of Physics, University of Adelaide, SA, Australia
}%
\author{Z.~Koumi}%
\affiliation{%
 CSSM, Department of Physics, University of Adelaide, SA, Australia
}%
\author{Y.~Nakamura}
\affiliation{RIKEN Center for Computational Science, Kobe, Hyogo 650-0047, Japan}
\author{H.~Perlt}
\affiliation{Institut für Theoretische Physik, Universit\"at Leipzig, 04109 Leipzig, Germany}
\author{P.E.L.~Rakow}
\affiliation{Theoretical Physics Division, Department of Mathematical Sciences,University of Liverpool, Liverpool L69 3BX, UK}
\author{G.~Schierholz}
\affiliation{Deutsches Elektronen-Synchrotron DESY, Notkestr. 85, 22607 Hamburg, Germany}
\author{H.~St\"uben}
\affiliation{Regionales Rechenzentrum, Universit\"at Hamburg, 20146 Hamburg, Germany}
\author{R.D.~Young}%
\affiliation{%
 CSSM, Department of Physics, University of Adelaide, SA, Australia
}%
\author{J.M.~Zanotti}
\affiliation{%
 CSSM, Department of Physics, University of Adelaide, SA, Australia
}%

\collaboration{CSSM/QCDSF/UKQCD Collaboration}
\noaffiliation

\date{\today}

\begin{abstract}
We present a lattice analysis of the light pseudoscalar mesons with consideration for the mixing between the flavour-neutral states $\pi^0$, $\eta$ and $\eta^\prime$. We extract the masses and flavour compositions of the pseudoscalar meson nonet in $n_f=1+1+1$ lattice QCD+QED around an SU(3)-flavour symmetric point, and observe flavour-symmetry features of the extracted data, along with preliminary extrapolation results for the flavour compositions at the physical point. A key result of this work is the observed mass splitting between the $\pi^0$ and $\eta$ on our ensembles, which is found to exhibit behaviour that is simply related to the corresponding flavour compositions. 
\end{abstract}

\maketitle

The quark flavour compositions of the $\pi^0$, $\eta$ and $\eta^\prime$ mesons are most familiar to us in the limit of exact SU(3)-flavour symmetry where the up, down and strange quarks are degenerate. In this limit the $\pi^0$ and $\eta$ belong to an octet, whilst the $\eta^\prime$ is pure flavour singlet. However, the breaking of SU(3)-flavour symmetry in nature permits the flavour compositions of these flavour-neutral (FN) pseudoscalar (PS) mesons to differ from their SU(3) octet-singlet forms. Understanding and quantifying this difference for the physical states is important for theoretical and phenomenological studies where interpolating operators are used to project onto the physical states. Furthermore, this type of \emph{mixing} is directly tied to our understanding of the extent of quark-flavour symmetry breaking in nature, as can be seen explicitly from $\chi$PT \cite{GASSER198277} or flavour-breaking \cite{Kordov:2019oer} expansions.

It is understood that the $\pi^0$ mixes weakly with the other FN pseudoscalars ($\mathcal{O}(1^\circ)$ \cite{Escribano:2020jdy}); an effect due solely to broken isospin symmetry, which is itself driven by differences in the up and down quark charges and masses. The mixing between the $\eta$ and $\eta^\prime$ is understood to be considerably larger (of the order $10$--$20^\circ$ \cite{Gan:2020aco}) and proceeds due to broken SU(3) flavour symmetry even in the isospin limit. The magnitude of the mixing between the $\pi^0$ and $\eta$/$\eta^\prime$, as well as the influence of broken isospin on the $\eta$--$\eta^\prime$ mixing, is yet to be determined.

Past lattice QCD studies \cite{Christ:2010dd, Dudek:2011tt, Michael:2013gka, Ottnad:2017bjt, Bali:2021qem} have worked in the isospin limit, hence excluding $\pi^0$ admixture, and presented results for the $\eta$--$\eta^\prime$ mixing with reasonable consensus and agreement with phenomenology \cite{Gan:2020aco}. It is important to note that the majority of these existing $\eta$--$\eta^\prime$ mixing studies, lattice and otherwise, have focused on the mixing of decay constants (defined through the couplings of the mesons to axial-vector currents), either under the assumption that their mixing behaviour is mirrored in the state mixing or without reference to the mixing of the states. The decay constant picture has many interesting aspects, not least of which being its proximity to the axial anomaly \cite{Witten:1978bc}, but it is understood that in general, the mixing of FN PS states through the coupling to pseudoscalar operators will not follow that observed in the decay constants \cite{Feldmann:1998sh}.

The FN mesons present a particular challenge to lattice QCD+QED in the calculation of their quark-loop contributions, which require determinations of self-to-self quark propagators. Direct calculation of the self-to-self propagator bears the same computational expense as the all-to-all propagator, which is prohibitively high, and necessitates methods of approximation which typically rely on the cancellation of introduced stochastic noise. Since the non-trivial $\pi^0$--$\eta$--$\eta^\prime$ mixing proceeds entirely through disconnected loop diagrams \cite{Christ:2010dd}, achieving a good self-to-self propagator signal while controlling computational cost is a necessity for this study. We address this difficulty using a combination of $\mathbb{Z}_2$-noise wall sources, dilution, and both source and sink gauge-covariant Gaussian smearing. 

In this work we extract the masses and flavour compositions of the PS mesons near an SU(3)-flavour symmetric point using $n_f=1+1+1$ lattice QCD+QED for the first time, including resolving the $\pi^0$--$\eta$ mass-splitting. We present and fit flavour-breaking expansions to our lattice results which are shown to perform well around the mass region where we have simulations, and present preliminary extrapolation results for the meson flavour compositions at the physical quark masses. We do not presently perform extrapolations to the continuum or infinite volume, which is instead reserved for future work.
\newpage

\section{\label{sec:method}Diagonalization on the lattice\protect\\}

To study the FN PS mesons on the lattice one must choose a set of interpolating operators which couple to them. If the up, down and strange quarks are degenerate, then the familiar SU(3) octet-singlet operators
\begin{equation*}
    \mathcal{O}_{\pi_3} = \frac{1}{\sqrt{2}}(\bar{u}\gamma^5u - \bar{d}\gamma^5d), \, \mathcal{O}_{\eta_8} = \frac{1}{\sqrt{6}}(\bar{u}\gamma^5u + \bar{d}\gamma^5d - 2\bar{s}\gamma^5s),
\end{equation*}
\begin{equation}
    \mathcal{O}_{\eta_1} = \frac{1}{\sqrt{3}}(\bar{u}\gamma^5u + \bar{d}\gamma^5d + \bar{s}\gamma^5s),
\end{equation}
here defined with isospin symmetry, couple diagonally to the FN PS mesons. However, if the quarks are no longer degenerate due to the inclusion of QED or non-degenerate bare masses, in general these operators will have non-trivial overlap with each of the FN PS states. 

In this work we have made the assumption that the set of states coupled to by the octet-singlet basis operators above, or some other set of operators (e.g. the quark-flavour basis) related by a simple change of basis, are a complete set of states with respect to the low-lying mass eigenstates $\pi^0$, $\eta$ and $\eta^\prime$. Additionally, for the lattice volumes and large quark masses used in this study we need not consider contamination by other low-lying states, such as $2\gamma$ and $3\pi$ channels, due to their relatively high energies. 
Although in principle there can be mixing between our states of interest and glueball or heavy quark operators, we expect our interpretation of the flavour compositions herein to be a good approximation of the low energy physics, as these additional states are understood to have negligible overlap with the FN PS mesons at our level of precision \cite{Ottnad:2012fv}.

\subsection{Operator basis and correlation functions}

We employ a variational basis of six interpolating operators; the three quark-flavour basis states 
\begin{equation}
  \mathcal{O}_u = \bar{u}\gamma^5 u, \quad \mathcal{O}_d = \bar{d}\gamma^5 d, \quad \mathcal{O}_s = \bar{s}\gamma^5 s, \label{eqn:operators}
\end{equation}
with two different levels of gauge-covariant Gaussian smearing each. Using these operators we construct a $6\times 6$ matrix of correlation functions (correlation matrix) with elements
\begin{equation}
    C_{ij}(t) = \sum_{\vec{x}, \vec{y}} \langle \mathcal{O}_j(\vec{y},t) \, \mathcal{O}^\dagger_i(\vec{x},0) \rangle, \label{eqn:correlators}
\end{equation}
where $i,j$ enumerate the six aforementioned interpolating operators.

The Wick contractions for the above correlation functions of two FN PS Dirac bilinears (Eq.~\ref{eqn:operators}) lead to two general combinations of quark propagator traces:
\begin{equation}
    C(t)_{disc} = \sum_{\vec{x},\vec{y}}\textrm{Tr}\left[S_{f}(\vec{y},t;\vec{y},t)\gamma^5\right]\textrm{Tr}\left[S_{f'}(\vec{x},0;\vec{x},0)\gamma^5\right],
\end{equation}
is a disconnected contribution corresponding to quark loops of flavours $f$ and $f'$, and
\begin{equation}
    C(t)_{con} = -\sum_{\vec{x},\vec{y}}\textrm{Tr}\left[S_{f}(\vec{y},t;\vec{x},0)\gamma^5S_{f}(\vec{x},0;\vec{y},t)\gamma^5\right],
\end{equation}
for a connected contribution from a quark flavour $f$. The traces are over both spin and colour degrees of freedom (DOF). Where the flavours of the source and sink operators differ, such as for the off-diagonal components of our correlation matrix, the corresponding correlation function is given by $C(t)_{disc}$. For source and sink operators of the same flavour, the diagonal components of our correlation matrix are given by $C(t)_{disc} + C(t)_{con}$.

\subsection{Stochastic wall source methods}

The quark propagators required in this study are calculated using stochastic $\mathbb{Z}_2$ noise sources with spin, colour and time dilution \cite{Foley:2005ac}. Dilution of a noise source in a particular DOF means that each wall source is separated into disjoint sources that are only non-zero for a single value of the diluted DOF, i.e. for spin, colour and time dilution we can write
\begin{equation}
    \eta_{r}(\vec{x},t;t_0)^{ab}_{\mu\nu} = \xi_{r}(\vec{x}) \, \delta_{tt_0} \, \delta_{ab} \, \delta_{\mu\nu},
\end{equation}
where Latin and Greek indices correspond to colour and spin degrees of freedom respectively. The spatial sources $\xi_{r}(\vec{x})$ are randomly generated from a uniform $\mathbb{Z}_2 \cong \{-1,1\}$ distribution, and hence exhibit the key property
\begin{equation}
    \lim_{N_r\rightarrow\infty}\frac{1}{N_r}\sum_{r=1}^{N_r} \xi_{r}(\vec{x}) \, \xi_{r}(\vec{y}) = \delta_{\vec{x}\vec{y}},
\end{equation}
where the index $r$ enumerates independently generated sources.

Using these diluted noise sources, the solution vectors are obtained by numerically solving
\begin{equation}
    \psi_{r}(\vec{y},t;t_0)^{ab}_{\mu\nu} = \sum_{\vec{z}} M^{-1}(\vec{y},t;\vec{z},t_0)^{ab}_{\mu\nu} \, \xi_{r}(\vec{z}). \label{eqn:solnvec}
\end{equation}
The all-to-all propagator can hence be approximated from an ensemble of $N_r$ independent noise sources as
\begin{equation}
    S(\vec{y},t;\vec{x},t_0)^{ab}_{\mu\nu} \approx \frac{1}{N_r} \sum_{r=1}^{N_r} \psi_{r}(\vec{y},t;t_0)^{ab}_{\mu\nu} \, \xi_{r}(\vec{x}),
\end{equation}
by computing a solution vector for each diluted source (i.e. each value of $t_0$, $b$ and $\nu$). The `self-to-self' quark propagator required for the calculation of $C(t)_{disc}$ is then recovered by setting $\vec{y}=\vec{x}$ and $t=t_0$.

We calculate the self-to-self propagator using the above method on each of the $N_t$ time-slices of the lattice (each value of $t_0$), which allows us to make a further improvement of the disconnected signal by averaging over the source times
\begin{equation*}
    C(\delta t)_{disc} = 
    \frac{1}{N_t}\sum_t \sum_{\vec{x},\vec{y}}\textrm{Tr}\left[S_{f}(\vec{y},t+\delta t;\vec{y},t+\delta t)\gamma^5\right] \dots 
\end{equation*}
\begin{equation}
    \times \textrm{Tr}\left[S_{f'}(\vec{x},t;\vec{x},t)\gamma^5\right].
\end{equation}

 The one-end trick is utilized in the calculation of the connected contributions to the correlation functions $C(t)_{con}$, where the solution vectors (Eq.~\ref{eqn:solnvec} for a single value of $t_0$) are used in place of the full propagators, granting a `free' sum over spatial source locations in the large $N_r$ limit (see, for example, \cite{Alexandrou:2008ru}). 
 
 Each correlation function is calculated using three noise sources on each configuration and $\mathcal{O}(1000)$ configurations on each ensemble. This configuration of diluted noise sources was found to deliver a sufficient signal for our operators, having tested various levels of noise reduction, however the number of inversions required per self-to-self propagator is $3\, {N}_\textrm{t}$ times that of a typical point-to-all propagator. Fortunately, different source smearings of the self-to-self propagator come at no additional cost since they are applied to the source after inversion and before construction of the full propagator (see, e.g. \cite{Bali:2021qem}).

\subsection{Diagonalization}

Since in this work we consider a total of six operators, we are able to resolve no more than six states in our simulations. Hence we assume that after a sufficient amount of time our correlation functions will receive contributions from the six lowest energy states in the system, and so at large times the elements of the correlation matrix, Eq.~\ref{eqn:correlators}, can be written as
\begin{equation}
    C_{ij}(t) = \sum_{n=1}^6 \langle 0 | \mathcal{O}_j | n \rangle \langle n | \mathcal{O}^\dagger_i | 0 \rangle e^{-M_n t}, \label{eqn:corrmatrix}
\end{equation}
where the states $|n\rangle$ are the mass eigenstates of the Hamiltonian, $|0\rangle$ is the physical vacuum, and there exist time-independent vectors $\vec{v}_m$ and $\vec{u}_m$ with the properties \cite{Blossier:2009kd} 
\begin{equation}\sum_j \langle 0 | \mathcal{O}_j | n \rangle v_{j,m} = \delta_{nm}, \quad \sum_j u_{j,m} \langle n | \mathcal{O}^\dagger_j | 0 \rangle = \delta_{nm}. \label{eqn:evecdelta}
\end{equation}
These vectors can be calculated as the solutions to the generalized eigenvalue problem (GEVP)
\begin{equation}
    C(t_0)^{-1} \, C(t_0+\delta t) \, \vec{v}_n = e^{-M_n\delta t} \, \vec{v}_n, 
\end{equation}
and similarly for the left eigenvectors $\vec{u}_n$. We solve this GEVP and diagonalize the correlation matrix at large times as $\vec{u}_n \, C(t) \, \vec{v}_n \propto e^{-M_n t}$, from which the masses $M_n$ can easily be determined. The three lowest energies correspond to the FN PS mesons of interest here, while the other three eigenvalues are discarded. Note that the correlation matrix is real and symmetric and hence the left and right generalized eigenvectors are each other's transpose. It is also noteworthy that the GEVP eigenvectors are also eigenvectors of the correlation matrix $C(t)$.

Once the masses have been determined we can extract the overlaps which encode the flavour compositions of the energy eigenstates by fitting
\begin{equation}
    e^{M_n t}\sum_{j=1}^6 C_{ij}(t)v_{j,n} = \langle n | \mathcal{O}^\dagger_i | 0 \rangle ,
\end{equation}
to a constant at sufficiently large $t$, for each $n \leq 3$ and $i$. 

We now wish to contrast the overlaps of operators with different amounts of quark smearing, and to that end we relabel our operators $\mathcal{O}_i \rightarrow \mathcal{O}_f^{(l)}$, where the index $f$ labels the quark flavour and $l$ the smearing level explicitly.

Given a fixed smearing level $l$, we identify the relative weight of flavour $f$ in eigenstate $|n\rangle$ by
\begin{equation}
     \langle 0 | \tilde{\mathcal{O}}^{(l)}_{f} | n \rangle \equiv \frac{\langle 0 | \mathcal{O}^{(l)}_f | n \rangle}{\sqrt{\sum_{f^\prime=u,d,s} |\langle 0 | \mathcal{O}^{(l)}_{f^\prime} | n \rangle|^2}}, \quad f=u,d,s. \label{eqn:rescaledolaps}
\end{equation}

In solving the GEVP we have chosen $\delta t = 1$, and the generalized eigenvectors are calculated at both timeslices $t_0=4$ and $t_0=5$, and the resulting overlaps and masses calculated from diagonalization by each eigenvector are averaged in order to capture some of the uncertainty associated with the choice of eigenvector.

\subsection{Lattice details}

\begin{figure}
\includegraphics[width=0.48\textwidth]{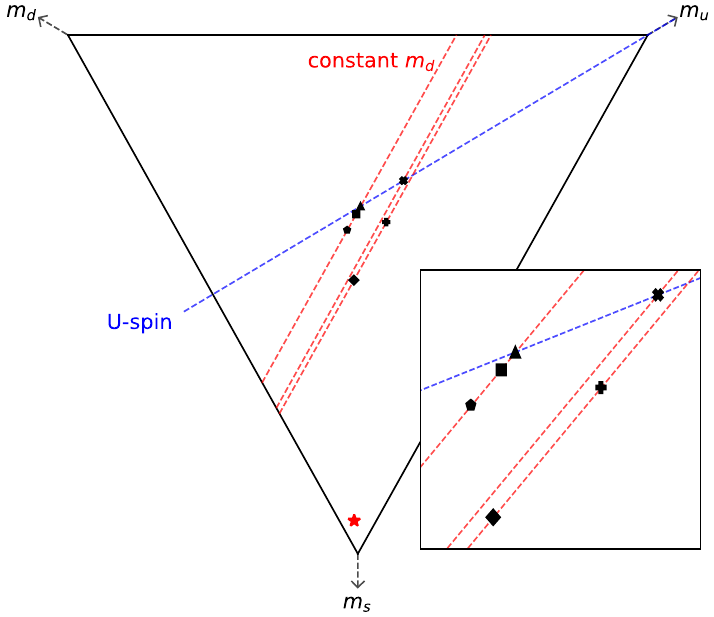}
\caption{A visualisation of our six ensembles on the plane of constant average bare quark mass, including a close-up view of the ensemble locations, along with the physical point indicated by a red star. The red dashed lines indicate paths of constant $m_d$, whilst the blue dashed line shows where the down and strange quarks are degenerate and thus U-spin symmetry is exact. The ensembles are 1 (triangle), 2 (square), 3 (pentagon), 4 (x), 5 (plus) and 6 (diamond). \label{fig:ensemblestriangle}}
\end{figure}

All correlation functions are calculated on $24^3\times48$, $n_f=1+1+1$, dynamical QCD+QED lattice gauge field ensembles around a U-spin symmetric point ($m_d=m_s$) with $m_u$ tuned to approximate SU(3) symmetry, as detailed in \cite{Horsley:2015vla}. These ensembles are confined to a plane of constant average (bare) quark mass, $\bar{m} = (m_u+m_d+m_s)/3 = m_0 = constant$. The quark hopping parameters and the extracted PS meson masses for each ensemble can be found in Table~\ref{tab:lattices}. Ensembles~1--3 were chosen to exhibit interesting mixing behaviour based on the approximate iso- (or T-), U- and V-spin symmetry observed in purely connected pseudoscalar meson masses along the $\delta m_d = 0$ trajectory. Ensembles~4--6 were generated secondarily in order to better constrain our parametrizations through a variation in the down quark mass. Our ensembles are depicted on the plane of constant $\bar{m}$ in Figure~\ref{fig:ensemblestriangle}, along with the physical point. Lines of constant $m_d$ which our ensembles lie on are denoted by the red dashed lines while the U-spin symmetric line is shown by the blue dashed line. 

The gauge actions used are the tree-level Symanzik improved SU(3) gauge action and the noncompact U(1) QED gauge action (further details in \cite{Horsley:2019wha, Horsley:2015vla,Horsley:2015eaa}). The fermions are described by an $\mathcal{O}(a)$-improved stout link non-perturbative clover (SLiNC) action \cite{Cundy:2009yy}. The couplings used and lattice spacing are 
\begin{equation}
\beta_{\textrm{QCD}} = 5.5, \quad \beta_{\textrm{QED}}=0.8, \quad a = 0.068(2) \, \textrm{fm},
\end{equation}
which gives a QED coupling $\alpha_\textrm{QED} \simeq 0.1$, roughly $10\times$ larger than the physical value.

\begin{table*}
\centering
\begin{tabular}{| p{.4cm} p{1.5cm} p{1.5cm} p{1.6cm}  p{1.4cm}  p{1.4cm}  p{1.4cm}  p{1.6cm}  p{1.4cm}  p{1.4cm}  p{1.4cm} |}
 \hline
 \multicolumn{11}{|c|}{Lattice ensembles and masses (MeV)} \\
 \hline
  \# & $\kappa_u$ & $\kappa_d$ & $\kappa_s$ & $M_{\pi^0}$ & $M_{\eta}$ & $M_{\eta}$-$M_{\pi^0}$ & $M_{\eta^\prime}$ & $M_{\pi^+}$ & $M_{K^+}$ & $M_{K^0}$ \\
  \hline
    1 & 0.124362 & 0.121713 & 0.121713 & 457(5) & 473(5) & 15(1) & 1234(51) & 485(4) & 485(4) & 459(5) \\
  \hline
    2 & 0.124374 & 0.121713 & 0.121701 & 475(4) & 483(4) & 10(1) & 1219(118) & 491(4) & 498(4) & 477(4) \\
  \hline
    3 & 0.124400 & 0.121713 & 0.121677 & 446(9) & 476(7) & 28(1) & 1165(54) & 461(8) & 478(8) & 474(7) \\
  \hline
    4 & 0.124281 & 0.121752 & 0.121752 & 430(11) & 521(9) & 86(3) & 1519(127) & 519(7) & 519(7) & 429(10) \\
  \hline
    5 & 0.124338 & 0.121760 & 0.121689 & 405(8) & 448(5) & 50(3) & 1156(70) & 437(5) & 464(4) & 413(6) \\
 \hline
    6 & 0.124430 & 0.121760 & 0.121601 & 404(7) & 503(5) & 97(2) & 1058(50) & 421(6) & 499(4) & 482(4) \\
 \hline

\end{tabular}
\caption{The ensemble number labels, $\kappa$ values and extracted PS masses for each of our six $n_f=1+1+1$ QCD+QED ensembles. \label{tab:lattices}}
\end{table*}

\begin{table*}
\centering
\begin{tabular}{| p{.4cm}  p{1.8cm}  p{1.8cm}  p{1.8cm}  p{1.7cm}  p{1.7cm} p{1.7cm}  p{1.7cm}  p{1.7cm}  p{1.7cm} |}
 \hline
 \multicolumn{10}{|c|}{Flavour compositions} \\
 \hline
  \# & $|\langle 0|\tilde{\mathcal{O}}_u| \pi^0 \rangle|^2$ & $|\langle 0|\tilde{\mathcal{O}}_d| \pi^0 \rangle|^2$ & $|\langle 0|\tilde{\mathcal{O}}_s| \pi^0 \rangle|^2$ & $|\langle 0|\tilde{\mathcal{O}}_u| \eta \rangle|^2$ & $|\langle 0|\tilde{\mathcal{O}}_d| \eta \rangle|^2$ & $|\langle 0|\tilde{\mathcal{O}}_s| \eta \rangle|^2$ & $|\langle 0|\tilde{\mathcal{O}}_u| \eta^\prime \rangle|^2$ & $|\langle 0|\tilde{\mathcal{O}}_d| \eta^\prime \rangle|^2$ & $|\langle 0|\tilde{\mathcal{O}}_s| \eta^\prime \rangle|^2$ \\
  \hline
    1 & 0.000(0) & 0.500(0) & 0.500(0) & 0.682(9) & 0.159(5) & 0.159(5) & 0.380(56) & 0.310(28) & 0.310(28) \\
  \hline
    2 & 0.111(46) & 0.641(12) & 0.248(53) & 0.590(49) & 0.019(12) & 0.391(58) & 0.380(26) & 0.312(14) & 0.309(13) \\
  \hline
    3 & 0.622(17) & 0.360(21) & 0.018(5) & 0.090(14) & 0.278(22) & 0.632(8) & 0.329(28) & 0.333(14) & 0.339(18) \\
  \hline
    4 & 0.000(0) & 0.500(0) & 0.500(0) & 0.656(18) & 0.172(9) & 0.172(9) & 0.462(46) & 0.269(23) & 0.269(23) \\
  \hline
    5 & 0.094(10) & 0.664(10) & 0.242(17) & 0.594(10) & 0.014(3) & 0.392(12) & 0.356(52) & 0.329(49) & 0.314(17) \\
 \hline
    6 & 0.511(20) & 0.488(19) & 0.001(1) & 0.225(36) & 0.161(25) & 0.614(27) & 0.337(46) & 0.278(41) & 0.385(67) \\
 \hline

\end{tabular}
\caption{The extracted overlaps squared of the physical states with the flavour basis operators on each of our ensembles. \label{tab:overlaps}}
\end{table*}

\begin{figure}
\includegraphics[width=0.48\textwidth]{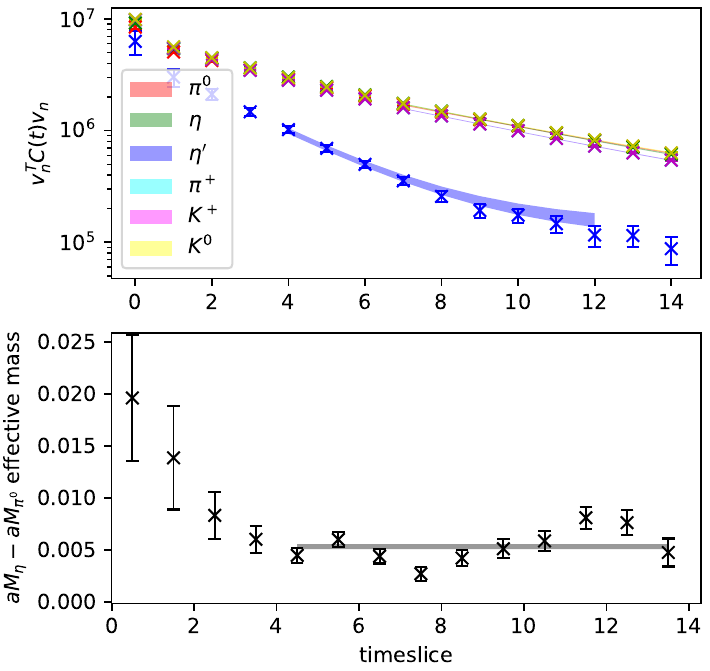}
\caption{An example of the diagonalized meson nonet correlation functions from our Ensemble~1 (top), as well as the effective mass difference of the $\pi^0$ and $\eta$ obtained from the ratio of their respective correlation functions (bottom). \label{fig:ens1effmass}}
\end{figure}

\section{\label{sec:results}Results and analysis\protect\\}

Our extrapolation scheme uses leading order (LO) flavour-breaking expansions in mass and charge parameters \cite{Bietenholz:2011qq,Horsley:2015vla} for the FN mesons (in the flavour basis) and the octet outer-ring:
\begin{widetext}

\begin{equation*}
    M^2_{FN} = \left[M^2_0 + \beta^\textrm{EM}_0(e_u^2+e_d^2+e_s^2)\right]\cdot\mathbb{I} + A \begin{bmatrix} 
        1 & 1 & 1 \\
        1 & 1 & 1 \\
        1 & 1 & 1 \\
    \end{bmatrix} + b_0\begin{bmatrix} 
        2\delta m_u & \delta m_u+\delta m_d & \delta m_u+\delta m_s \\
        \delta m_u+\delta m_d & 2\delta m_d & \delta m_d+\delta m_s \\
        \delta m_u+\delta m_s & \delta m_d+\delta m_s & 2\delta m_s \\
    \end{bmatrix} 
\end{equation*}
\begin{equation}
    + \, 2\beta_1^\textrm{EM}\begin{bmatrix} 
        e_u^2 & 0 & 0 \\
        0 & e_d^2 & 0 \\
        0 & 0 & e_s^2 \\
    \end{bmatrix} + \, 2\alpha \begin{bmatrix} 
        \delta m_u & 0 & 0 \\
        0 & \delta m_d & 0 \\
        0 & 0 & \delta m_s \\
        \end{bmatrix} 
    + a_1^\textrm{EM}\begin{bmatrix} 
        e_u^2 & e_u e_d & e_u e_s \\
        e_u e_d & e_d^2 & e_d e_s \\
        e_u e_s & e_d e_s & e_s^2 \\
    \end{bmatrix}, \label{eqn:FNmassexp}
\end{equation}
\end{widetext}
\begin{equation*}
    M^2_{\pi^+} = M^2_0 + \beta^\textrm{EM}_0(e_u^2+e_d^2+e_s^2) + \alpha(\delta m_u + \delta m_d)
\end{equation*}
\begin{equation}
    \, + \beta^\textrm{EM}_2(e_u-e_d)^2, \label{eqn:pi+exp}
\end{equation}
\begin{equation*}
    M^2_{K^+} = M^2_0 + \beta^\textrm{EM}_0(e_u^2+e_d^2+e_s^2) + \alpha(\delta m_u+\delta m_s) 
\end{equation*}
\begin{equation}
    \, + \beta^\textrm{EM}_2(e_u-e_s)^2, \label{eqn:k+exp}
\end{equation}
\begin{equation}
    M^2_{K^0} = M^2_0 + \beta^\textrm{EM}_0(e_u^2+e_d^2+e_s^2) + \alpha(\delta m_d+\delta m_s) , \label{eqn:k0exp}
\end{equation}
where the mass parameters $\delta m_i=m_i - m_0$ denote the deviation of the $i^\textrm{th}$ quark flavour's mass from its SU(3) symmetric starting point mass. In \cite{Horsley:2015vla} we considered an $8\times 8$ mass matrix for the PS meson octet which we herein extend to include the flavour singlet. We now have a $9\times 9$
 mass matrix, with a $3\times 3$ block for the FN mesons. In addition to the terms in 
 $8 \otimes 8$ considered in \cite{Horsley:2015vla}, we also have terms with the 
 symmetries $1 \otimes 1$, $1 \otimes 8$ and $8 \otimes 1$, 
 which are (trivially) decomposed as
 \begin{equation}
 1 \otimes 1 = 1, \quad
 1 \otimes 8 = 8, \quad
 8 \otimes 1 = 8.
 \end{equation}
 The $ 1 \otimes 1 $ term gives the $A$ term of Eq.~\ref{eqn:FNmassexp}, whilst the $1 \otimes 8 $ and $8 \otimes 1$ terms give 
 the $b_0$ term. Alternatively, in the language of the expansion employed in \cite{Kordov:2019oer}, the $A$ and $b_0$ terms arise as zeroth and first order mass terms in the Taylor expansion of the disconnected components of the FN correlation functions.

The fact our ensembles share a constant average quark mass implies that $\delta \bar{m} = 0$. Since we are not varying the quark charges in our ensembles, we have no way of constraining the relative magnitudes of the $M_0$ and $\beta_0^\textrm{EM}$ terms in our expansions, and they are hence absorbed into one term. As a result of this we cannot fully distinguish distinct QCD and QED contributions to our mass expansions.

The masses of the eigenstates $\pi^0$, $\eta$ and $\eta^\prime$, presented in Table~\ref{tab:lattices}, are fit to the eigenvalues of the matrix expansion Eq.~\ref{eqn:FNmassexp}, while the outer ring PS mesons are simultaneously fit to Equations~\ref{eqn:pi+exp}--\ref{eqn:k0exp}. We label the diagonalized states by their mass ordering, consistent with the orderings appearing in the physical spectrum rather than their flavour content. 

Also included in the fits are determinations of the mass difference $M_{\eta}-M_{\pi^0}$, obtained from effective mass fits of the ratio of the relevant correlation functions calculated from the GEVP diagonalization. An example of the diagonalized PS meson correlation functions and the $M_{\eta}-M_{\pi^0}$ effective mass for Ensemble~1 can be seen in Figure~\ref{fig:ens1effmass}, where we observe a clear $\pi^0$--$\eta$ mass splitting of $14(2) \, \textrm{MeV}$.

In order to fit the normalized overlaps $\langle 0 | \tilde{\mathcal{O}}^{(l)}_{i} | n \rangle$ (Eq.~\ref{eqn:rescaledolaps}) we notice that the relative couplings of each flavour to a given eigenstate does not depend significantly on the smearing for our ensembles. As an example of this, for Ensemble~1, we compare relative up and down operator overlaps with the $\eta$ for each smearing:
\begin{equation}
    \frac{\langle 0 | {\mathcal{O}}^{(1)}_{u} | \eta \rangle}{\langle 0 | {\mathcal{O}}^{(1)}_{d} | \eta \rangle} = -2.097(23), \quad \frac{\langle 0 | {\mathcal{O}}^{(2)}_{u} | \eta \rangle}{\langle 0 | {\mathcal{O}}^{(2)}_{d} | \eta \rangle} = -2.080(25), \label{eqn:smearingcomp}
\end{equation}
which clearly agree within their uncertainties. This would suggest that the operator diagonalization is largely selecting the flavour composition of the three low-lying states. In principle, the excited states could have a flavour composition that differs from the lowest-lying states, but we would be unable to resolve these features in the present analysis. While we cannot yet conclude anything about the excited states, the two levels of smearing do help to improve the ground state isolation at early times. Since flavour composition of the lowest states is essentially independent of the smearing, we drop the explicit smearing index in the following discussion.

Collapsing the smearing degree of freedom in our correlation matrix and Taylor expanding about an SU(3) symmetric point yields a parametrization with the same functional form as Eq.~\ref{eqn:FNmassexp}, and we hence fit the overlaps $\langle 0 | \tilde{\mathcal{O}}_{i} | n \rangle$, presented in Table~\ref{tab:overlaps}, to the eigenvectors of an expansion of this form. Since we are here treating the state mixing as being distinct from the mixing of the mass matrix, we do not fit the masses and overlaps simultaneously (i.e. the fit parameters need not take the same values).

\begin{figure*}
    \centering
    \includegraphics[width=0.95\textwidth]{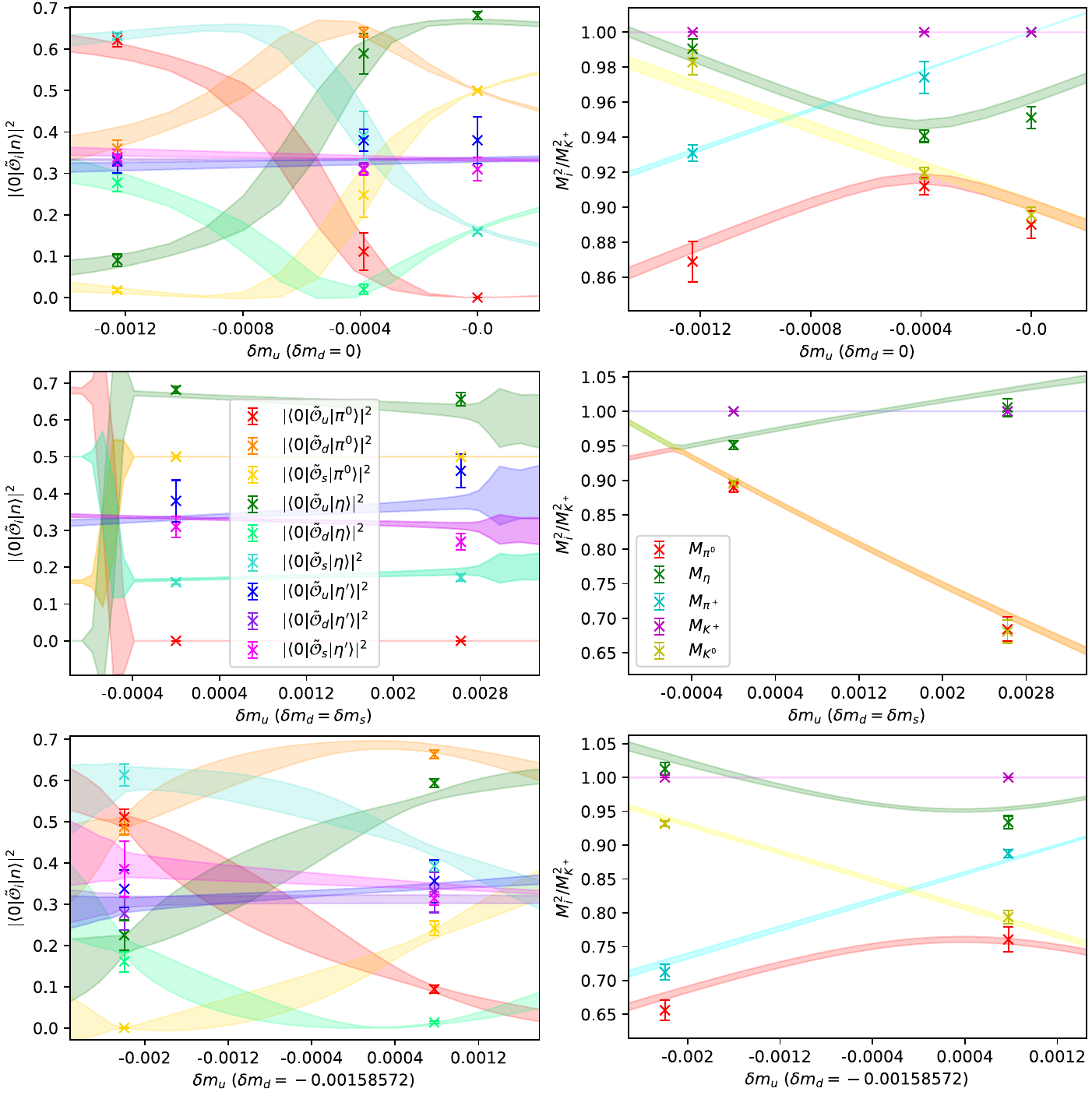}
    \caption{The overlaps (left column) and octet masses (right column) from each of our 6 ensembles with their respective global fits. The top pair of plots display Ensembles~~1--3 (right-to-left), which lie on the constant down quark mass trajectory $\delta m_d =0$ (left-most dashed red line of Fig.~\ref{fig:ensemblestriangle}). The center plots depict Ensembles~1 and 4, which both exhibit U-spin symmetry (dashed blue line in Fig~\ref{fig:ensemblestriangle}), whilst the bottom pair of plots depict Ensembles~5 and 6, which lie on the constant down quark mass trajectory $\delta m_d = -0.00159$ (right-most dashed red line of Fig.~\ref{fig:ensemblestriangle}). For a complete discussion of the features of this figure refer to Section~\ref{sec:results}.}
    \label{fig:allMassandOlaps}
\end{figure*}

In Figure~\ref{fig:allMassandOlaps} we present a visualization of the overlaps and octet masses from each of our 6 ensembles, along with their respective global fits. The $\chi^2/\textrm{dof}$ for the two global fits are 2.4 (overlaps) and 2.1 (masses). Whilst these $\chi^2/\textrm{dof}$ values are large, for the present preliminary investigation we deem the fits acceptable for the purpose of commenting on some general features of the extracted data. Moreover, given the small lattice volumes and additional systematic uncertainties present, the quoted statistical uncertainties for our lattice data likely underestimate the true uncertainties. Simply adding an additional 2\% uncertainty to our data as a conservative estimate of the systematic uncertainties lowers the $\chi^2/\textrm{dof}$ values to 1.1 and 0.6 for the overlap and mass fits respectively.

The top row of the plots depicts ensembles 1--3 (right-to-left), which all lie on the constant down quark mass trajectory $\delta m_d = 0$. Ensemble~~1 exhibits U-spin symmetry due to the degeneracy of the down and strange quarks, and we can see that the lightest FN PS meson exhibits the exact state composition of a U-spin $\pi_3$, $\pi_3^U = (\bar{d}\gamma^5d - \bar{s}\gamma^5 s)/\sqrt{2}$, with its mass necessarily degenerate with that of the $K^0$. The $\eta$ meson of Ensemble~2 has a state composition approaching that of a V-spin $\pi_3$, $\pi_3^V = (\bar{u}\gamma^5u - \bar{s}\gamma^5 s)/\sqrt{2}$. Ensemble~2 also appears to be very near the waist of an avoided level crossing between the $\pi^0$ and $\eta$ mesons. Between Ensembles~2 and 3, as determined from the overlap fit, the $\pi^0$ becomes a pure isospin $\pi_3$ at $\delta m_u \approx -0.0008$.

The two plots that occupy the center row of Figure~\ref{fig:allMassandOlaps} show the overlaps and octet masses of Ensembles~1 and 4 (left-to-right), which lie along a quark mass trajectory where the down and strange quarks have equal masses, as fixed by our condition $\delta \bar{m}=0$. Along this trajectory we have hence enforced U-spin symmetry, and one of either the $\pi^0$ or $\eta$ exhibit the flavour structure of a $\pi_3^U$ throughout. A distinct feature of these plots is the level crossing observed in the $\pi^0$ and $\eta$ masses at $\delta m_u \approx -0.0004$, and corresponding point in the overlaps where the state compositions change labels according to the mass ordering. Additionally, one can observe that all three FN PS states approach their SU(3)-symmetric flavour compositions at approximately the location of Ensemble~1. The separation of the level crossing and exact SU(3)-flavour-states points is an EM effect, as without EM these two phenomena would always occur together at points with equal light quark masses (i.e. exact SU(3)-flavour symmetry). Additionally the mass splitting of about $2.5 \%$ between the charged and neutral octet mesons at the point of $M_{\pi^0}$--$M_\eta$ degeneracy is also a pure EM effect.

The plots occupying the bottom row of Figure~\ref{fig:allMassandOlaps} depict the overlaps and octet masses of Ensembles~5 and 6 (right-to-left), which are situated on our lightest constant down quark mass trajectory, $\delta m_d = -0.0016$. The flavour composition of the $\eta$ of Ensemble~5 can be seen to be near that of a $\pi_3^V$, whilst the $\pi^0$ of Ensemble~6 is a very good approximation of an isospin $\pi_3$. Ensemble~6 exhibits the poorest overlap signal in our set, likely since it also possesses the lightest up and down quarks. The octet masses again exhibit an avoided level crossing between the $\pi^0$ and $\eta$, however with a much broader waist than that observed around Ensemble~2.

\begin{figure}
\includegraphics[width=0.48\textwidth]{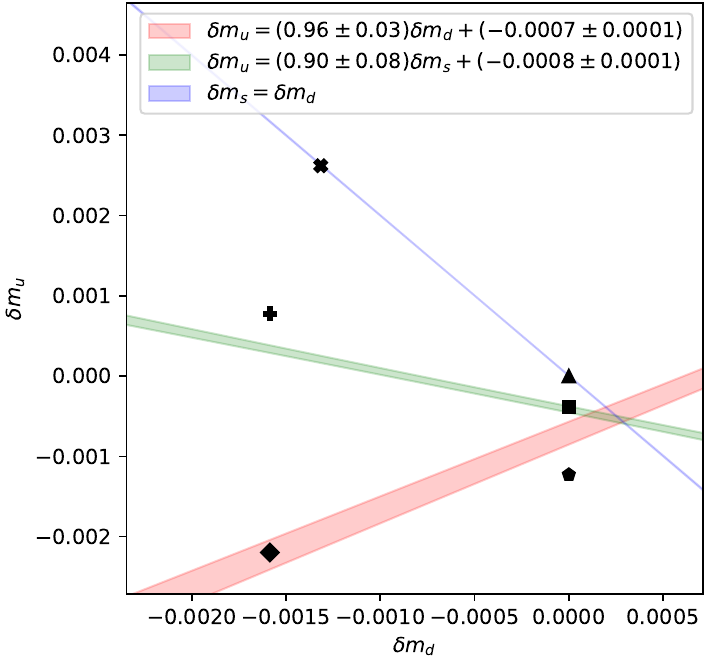}
\caption{Lines of pure isospin (red), U-spin (blue) and V-spin (green) $\pi_3$ states as determined from our global fit to the extracted overlaps. The linear relationships in the legend indicate the symmetry condition between the relevant pairs of quarks, which are non-trivial for isospin and V-spin due to the presence of EM. Also illustrated are the locations of our Ensembles 1 (triangle), 2 (square), 3 (pentagon), 4 (x), 5 (plus) and 6 (diamond). }
\label{fig:symfitlines}
\end{figure}

It is interesting to note that across the range of quark masses considered, the state compositions evolve between the distinct SU(2) subgroups: T-spin, U-spin and V-spin. In particular, as highlighted above, as $\delta m_u$ changes in the top-left panel of Figure~\ref{fig:allMassandOlaps}, we observe three distinct locations where one of the eigenstates appears as a pure $\pi_3$ state of a distinct SU(2) subgroup. Similarly, the lower-left panel also identifies pure $\pi_3^T$ and $\pi_3^V$ at particular values of $\delta m_u$. Using the parametrized description of the state composition we can trace out these distinct SU(2) subgroups in the quark mass plane, as shown in Figure~\ref{fig:symfitlines}. Each of the three lines corresponds to a trajectory where one of the eigenstates is a pure $\pi_3$. While the U-spin trajectory is exact, the isospin and V-spin trajectories have slopes that are roughly compatible with maintaining degeneracy of the quark masses.

Figure~\ref{fig:symfitlines} also suggests an improved definition of the SU(3) symmetric point, where the three lines appear to intersect together at a down quark mass slightly heavier than that of the nominal SU(3) symmetric point. This intersection point also coincides with the point where the $\pi^0$ and $\eta$ are degenerate along the U-spin symmetric line. While the nominal symmetric point was chosen such that the connected-only flavour neutrals are degenerate \cite{Horsley:2019wha,Horsley:2015eaa}, the location identified here uses only physical states in the spectrum. In practice however, tuning lattice quark masses with respect to the disconnected correlation functions needed in this study would be unfeasible, and from the point of view of an expansion about an approximate SU(3) symmetric point, the consequence for any physical observable will always be equivalent up to the order of an expansion.

Although we currently lack ensembles at large enough $|\delta m_i|$ to effectively resolve the physical point mixing, we can assess our overlap extrapolation at the physical values of the quark masses, $\delta m_i^*$, the locations of which were determined in \cite{Horsley:2015vla}, albeit on a $32^3\times64$ volume. We note that in this preliminary work we make no attempt to quantify the finite volume or lattice spacing effects in our results. We scale the parameters in our expansion that arise due to EM (note that $\beta_0^\textrm{EM}$ doesn't contribute to the mixing) as was done in \cite{Kordov:2019oer} to approximately correct our larger-than-physical EM coupling, and find
\begin{equation}
    | \pi^0  \rangle  = 0.85(14)|\pi_3 \rangle \, - \, 0.27(25) |\eta_8 \rangle \, + \, 0.29(22)|\eta_1 \rangle, \label{eqn:pioverlapextrap}
\end{equation}
\begin{equation}
    | \eta  \rangle  = -0.07(10)|\pi_3 \rangle \, + \, 0.76(16) |\eta_8 \rangle \, + \, 0.56(24)|\eta_1 \rangle,
\end{equation}
\begin{equation}
    | \eta^\prime  \rangle  = -0.005(2)|\pi_3 \rangle \, - \, 0.26(10) |\eta_8 \rangle \, + \, 0.96(3)|\eta_1 \rangle. \label{eqn:etapoverlapextrap}
\end{equation}
With our relatively low level of precision at the physical point we cannot resolve much significant mixture of the $\pi^0$ with either the $\eta$ or $\eta^\prime$, but we can see a small non-zero $\pi_3$ content in the $\eta^\prime$, although also too early to draw any physical conclusions. We also observe some non-trivial admixtures of the $\eta_8$ and $\eta_1$ occuring in the physical $\eta$ and $\eta^\prime$, and since all four numbers are consistent with a parametrization by a single mixing angle, we present a determination of said angle as
\begin{equation}
    \theta_{\eta\eta^\prime}=\sin^{-1}(-0.26\pm 0.10) = (-15.1^{+5.9}_{-6})^\circ,
\end{equation}
which is consistent with existing results from lattice QCD \cite{Christ:2010dd, Dudek:2011tt, Ottnad:2017bjt} and phenomenology \cite{Bramon:1997va, Feldmann:1998vh}.


While the extrapolation of the mixing angles to the physical point is largely exploratory, we use the masses as a benchmark to quantify the limitations of the present extrapolation. Using the physical quark mass point from \cite{Horsley:2015vla}, as above, we determine physical meson masses that are within 10-15\% of observation. For instance, on the present small volume and low-order chiral extrapolation we obtain $M_{\pi^+} = 114(17) \textrm{MeV}$ and $M_{K^+} = 551(2) \textrm{MeV}$. If however we choose to constrain the quark mass parameters to give the physical $M_{\pi^+}$ and $M_{K^+}$, the mixing angles do not appreciably differ from those reported above. As an example, with the retuned quark masses, the $\eta$--$\eta^\prime$ mixing is determined to be $\theta_{\eta \eta^\prime}=(-12.9^{+7.5}_{-7.8})^\circ$, in agreement with the number reported above.

\section{\label{sec:conclusion}Conclusion and outlook\protect\\}

In this investigation we have, for the first time, computed and resolved the broken-isospin induced $\pi^0$--$\eta$ mass-splitting near an effective SU(3) symmetric point, as well as observed the qualitative effects of electromagnetism (EM) and broken isospin on the flavour compositions of the flavour-neutral (FN) pseudoscalar (PS) mesons.

We have shown the efficacy of studying the FN PS mesons through the use of stochastic noise sources in combination with gauge-covariant Gaussian smearing and the variational method. Further, we have presented what appears to be a promising method for studying the overlaps of the FN PS mesons with respect to the chosen interpolating operator basis, and shown that they can be further understood and extrapolated by appropriate parametrization. It is clear that an understanding of the mixing behaviour of the FN states sheds light on the corresponding masses.

The results of this study give us confidence that a future analysis following these methods and utilizing larger lattice volumes and physical QED coupling should reproduce the PS meson masses accurately, as well as resolve the EM and isospin breaking effects on the flavour contents of the FN PS mesons at physical pion mass.

\begin{acknowledgments}
We would like to thank A. Hannaford-Gunn for their helpful feedback and discussion during the preparation of this manuscript. The numerical configuration generation (using the BQCD lattice QCD program \cite{Haar:2017ubh} with single quark flavours treated in the HMC by the tRHMC algorithm \cite{Haar:2018jjd}) and data analysis (using the Chroma software library \cite{Edwards:2004sx} and a GPU-accelerated mixed-precision conjugate gradient fermion matrix inverter through the COLA software \cite{Kamleh:2012sh}) was carried out on the DiRAC Blue Gene Q and Extreme Scaling (EPCC, Edinburgh, UK) and Data Intensive (Cambridge, UK) services, the GCS supercomputers JUQUEEN and JUWELS (NIC, Jülich, Germany) and resources provided by HLRN (The North-German Supercomputer Alliance), the NCI National Facility in Canberra, Australia (supported by the Australian Commonwealth Government) and the Phoenix HPC service (University of Adelaide). ZRK was supported by an Australian Government Research Training Program (RTP) Scholarship. 
RH was supported by STFC through grant ST/P000630/1.
WK was supported by Australian Research Council Grants DP19012215, DP210103706 and the Pawsey Centre for Extreme Scale Readiness.
HP was supported by DFG Grant No. PE 2792/2-1.
PELR was supported in part by the STFC under contract ST/G00062X/1.
GS was supported by DFG Grant No. SCHI 179/8-1.
RDY and JMZ were supported by the Australian Research Council Grants
FT120100821, FT100100005, DP140103067 and DP190100297.
We thank all funding agencies.
\end{acknowledgments}

\bibliography{npsmixing}

\end{document}